\documentclass[12pt,a4paper]{article}
\usepackage{amsbsy}
\usepackage{amsmath}
\usepackage{graphicx}
\usepackage{caption}
\usepackage{subcaption}

\title{Series Solutions of ${\cal PT}$-Symmetric Schr\"odinger Equations}

\author{
C. Ford\thanks{c.ford@imperial.ac.uk }~~~ and~~~ B. Xia\thanks{bichang.xia13@imperial.ac.uk
}
\\
Department of Mathematics\\
 Imperial College London\\
London SW7 2AZ}

\begin{document}\maketitle

\begin{abstract}

We consider series  solutions of the Schr\"odinger equation for the Bender-Boettcher
potentials $V(x)=-(ix)^N$ with integer $N$.  A simple truncation is introduced which provides
accurate results regarding the ground state and  first few excited states for any $N$.
This is illustrated with explicit computations of energy levels, node structure and expectation values
for some integer $N$. 

\end{abstract}


\bigskip

The complex Schr\"odinger equation
\begin{equation}-\psi''(z)-(iz)^N\psi(z)=E\psi(z)\end{equation}
is known to have  positive spectra if $N\geq 2$ \cite{Bender:1998ke,Bender:1998gh,Dorey:2001uw}.
 If $N$ is integer the wave functions are entire functions and the complex plane splits naturally into $N+2$ Stokes wedges. 
Energy quantisation is a consequence of demanding that $\psi(z)$ decays exponentially in a ${\cal PT}$-symmetric pair of Stokes wedges. 
 For any energy $E$, real or complex, there is a solution which decays
exponentially in any given wedge.
  For special values of $E$ one can find solutions that decay in two
(non-contiguous) wedges.  

Consider the double power series
\begin{equation}\label{series1}
\psi_1(z)=\sum_{p=0}^\infty\sum_{q=0}^\infty a_{pq}~ (iz)^{(N+2)p+2q} E^q.
\end{equation}
where the $a_{pq}$ are constants. 
Inserting this into the Schr\"odinger equation yields
the recursion relation
\begin{equation}
[(N+2)p+2q-1]\cdot[(N+2)p+2q]~
a_{pq}=a_{p-1~q}+a_{p~q-1}.
\label{recursion1}
\end{equation}
Viewing $a_{pq}$ as a matrix, equation (\ref{recursion1})
expresses any element $a_{p q}$ in terms of the elements  directly above and 
directly to the left.
On fixing the top left element $a_{00}$ one can, in principle, determine all the other elements.
For the convenient choice $a_{00}=1$ all the $a_{pq}$ are positive rational numbers.
 With this choice
$\psi_1(0)=1$ and $\psi'_1(0)=0$.

A second solution of the Schr\"odinger equation is
\begin{equation}\psi_2(z)=\sum_{p=0}^\infty\sum_{q=0}^\infty b_{pq}~ (iz)^{1+(N+2)p+2q} E^q
\label{series2}\end{equation}
where the coefficients $b_{pq}$ satisfy
\begin{equation}
[(N+2)p+2q]\cdot[(N+2)p+2q+1]~
b_{pq}=b_{p-1~q}+b_{p~q-1}.
\label{recursion2}
\end{equation}
It is convenient to take $b_{00}=1$ so that  $\psi_2(0)=0$ and  $\psi_2'(0)=i$.

Consider a linear combination of  the 
two solutions\begin{equation}
\psi(z)=\psi_1(z)+c~\psi_2(z),\end{equation}
where $c$ is a complex constant.
By a suitable choice of $c$ one can ensure that $\psi(z)$ decays exponentially in any one
of  the $N+2$ Stokes wedges. For example, to obtain decay in the wedge centred
at (the anti-Stokes line) $\theta=-{1\over2}\pi(N-2)/(N+2)$  take
\begin{equation}c=-\left.\lim_{r\rightarrow\infty}{\psi_1(re^{i\theta})\over{\psi_2(r e^{i\theta})}}\right|_{
\theta=-{1\over2}\pi(N-2)/(N+2)}.\label{c}\end{equation}
Although this works for any $E$
it only gives a decaying wave function in one of the $N+2$ wedges. However, if 
both $E$ and $c$ are real
 the solution will also decay in the ${\cal PT}$
image of the  wedge. To determine the spectrum in a  ${\cal PT}$ symmetric pair of wedges
 it suffices to determine the real energies for which $c$ is real. This can be implemented graphically
 by plotting $\hbox{Im } c$ as a function of $E$ - the roots are the energy levels.
 
 \begin{figure}[ht]\centering
\includegraphics[width=4in]{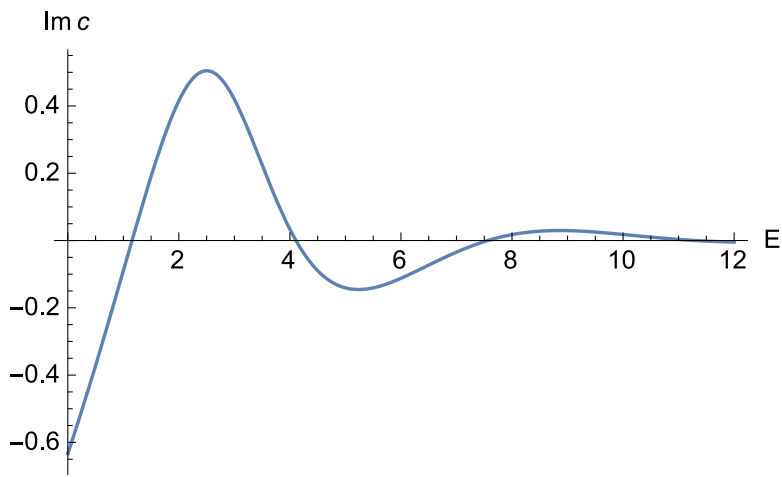}
\caption{ The first few energy levels in the $N=3$ theory  appear as roots of $\hbox{Im } c$}
\end{figure}

 In figure 1 this is plotted for $N=3$; as $E$ is increased $\hbox{Im } c$  approaches zero
 in an oscillatory fashion. 
 $\hbox{Im }c$ has no roots for negative $E$. To produce this plot we made two approximations:
 
 \smallskip
 \noindent i) In the double power series (\ref{series1}) and (\ref{series2}) we retained all terms with
 $p+q\leq 100$.
 
 \smallskip
\noindent ii) In equation (\ref{c}) a `large' finite value of $r$ is taken instead of the $r\rightarrow\infty$ limit.
 We took $r=8$.

\smallskip

  In the matrix language the truncation is anti-diagonal in that 
 entries below the $p+q=100$ line are discarded.
 Applying a root finding algorithm to our approximation for $\hbox{Im }c(E)$ one can compute the energy levels and associated real $c$ values.

 \begin{table}[ht]
\caption{Energy Levels and  $c$ Values in the $N=3$ Theory}
\centering
\begin{tabular}{c c c c}
\hline\hline
$n$ & $E_n$ & $c_n$ & \\ % inserts table %heading
\hline
0&  1.1562670719881132937   &-0.53871550451988192490\\
1&  4.1092287528096515358&-2.32727424075874334001& \\
2&    7.5622738549788280413  &-2.69833514190279036708 \\
3 &11.314421820195804397 &  -3.37823419494258452822\\
4&
15.291553750392532~~~~~
&-3.90980926012776641~~~~~
 \\
5&19.451529130691~~~~~~~~~~
  &-4.41178037226863~~~~~~~~~~
  \\
6&23.766740435~~~~~~~~~~~~~~
 & -4.87570168194~~~~~~~~~~~~~~~
    \\
7& 28.2175249~~~~~~~~~~~~~~~~~ 
&-5.312499663~~~~~~~~~~~~~~~~~~
      \\
\hline
\end{tabular}
\label{N=3values}
\end{table} 
\vfill\eject

 For large values of $n$,  $c_n$ is approximately
 $-\sqrt{E_n}$.
 To investigate the accuracy of our method one can vary the $p+q\leq 100$ truncation and the $r$ value. The values of the first few energy levels is not affected by taking $r=7$ instead of $r=8$ at least to 20 significant figures. Similarly, upping the truncation to $p+q\leq 150$ does not change the first few
 energy levels (again to 20 significant figures). However the higher energy levels are sensitive
 to changes in $r$ and the truncation. We have quoted $E_4$ to  17 rather than 20 significant figures
 as the missing three digits change when the truncation is improved. 
 For higher $n$ the accuracy drops further. As the double power series are expansions in $z$ and $E$ it is expected that the truncation is less accurate for higher energies.  Our  energy levels are consistent with the Runge-Kutta based results
    reported by Bender in \cite{Bender:2007nj}.

For higher $N$ there is more than one pair of (non-adjacent) ${\cal PT}$-symmetric wedges
 \cite{Schmidt:2012hj}. If $N$ is odd there are ${1\over2}(N-1)$ such pairs.
 If $N$ is even there are ${1\over2}(N+2)$ pairs; one of these pairs
 is also ${\cal P}$ symmetric where our graphical method is not applicable.\footnote{
 Here the energy eigenstates are of the form $\psi_1$ (even)
 or
 $\psi_2$  (odd).}
 For example,
there are three ${\cal PT}$-symmetric pairs if $N=7$. Each pair gives a distinct real and positive spectrum;
$\hbox{Im } c$ is plotted as a function of $E$ for the three pairs below.

\begin{figure}	
	\centering
	\begin{subfigure}[t]{3in}
		\centering
		\includegraphics[width=3in]{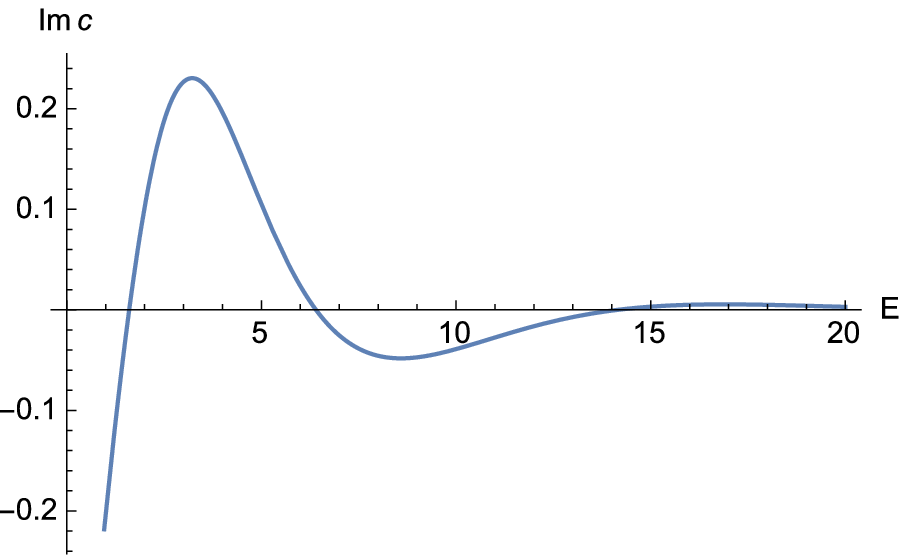}\medskip
			
	\end{subfigure}

	\begin{subfigure}[t]{3in}
		\centering
		\includegraphics[width=3in]{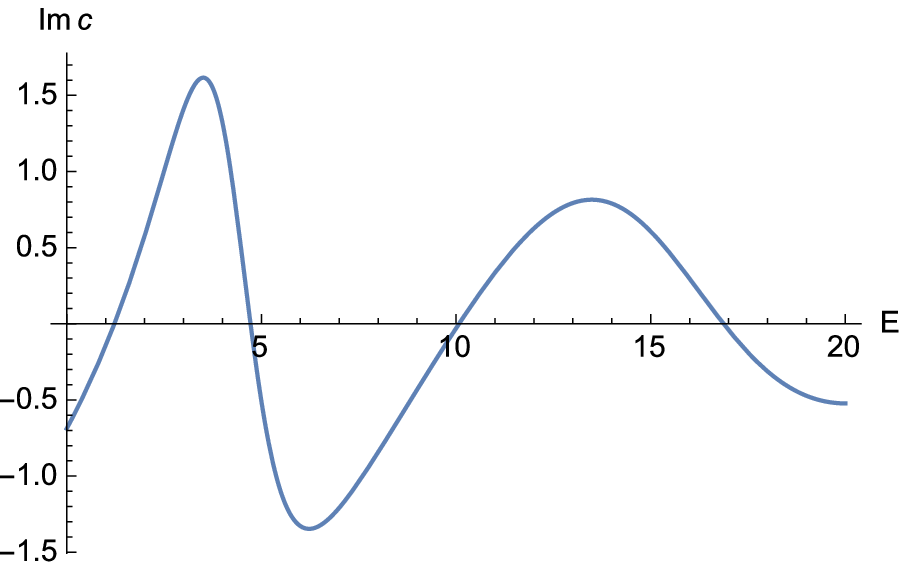}\medskip
	
	\end{subfigure}

	\begin{subfigure}[t]{3in}
		\centering
		\includegraphics[width=3in]{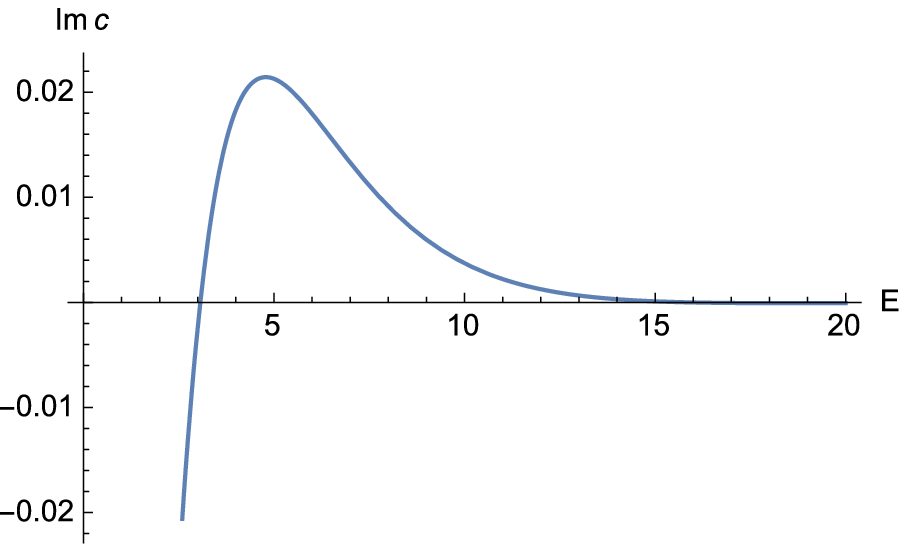}
		
	\end{subfigure}	
	\caption{$\hbox{Im }c$ in the $N=7$ theory for the three ${\cal PT}$-symmetric spectra.
	The upper plot is for the wedges centred at $\theta=\pi/6$ and $\theta=5\pi/6$,
	the middle plot has wedges centred at $\theta=-\pi/18$ and $\theta=-17\pi/18$
	and the lower plot has wedges centred at $\theta=-5\pi/18$ and $\theta=-13\pi/18$.}
	
	\label{fig:2}
\end{figure}

\vfill\eject

For higher $n$ the $E_n$ have ratios $1.41 ~:~1~:~3.52$ \cite{Schmidt:2012hj}.
Although our method ia adapted to small $n$ such behaviour is evident in the third
excited state; $E_3$ has values $23.702$, $16.872$, $59.026$.
Note that the ratios of the ground state energies are different;
$E_0$ has values $1.6047$, $1.2247$, $3.0686$.

For the `upper' spectrum (with wedges centred at $\theta=\pi/6$ and $\theta=5\pi/6$)
the $c_n$ values are positive with   $c_n\approx \sqrt{E_n}$ for large $n$.
The other two spectra yield negative $c_n$ with $c_n\approx -\sqrt{E_n}$.
The plots were produced via the same $p+q\leq 100$ truncation but with $r=3$
rather than $r=8$.\footnote{ For $N=7$ the truncation breaks down for lower $|z|$.}
Similar results can be obtained for higher $N$. For example the $N=19$ model
has 9 distinct spectra (4 giving positive $c_n$, 5 with negative $c_n$)

The truncations considered here can be used to identify the nodes of the energy eigenstates.
Although our truncation fails for large enough $|z|$, at least for the first few energy levels,
the nodes are close enough to the origin for them to be located with high precision. 
Returning to the $N=3$ case all energy eigenfunctions have an infinite string
of zeros on the positive imaginary axis; for each energy level these are above
the classical turning point $iE_n^{1/3}$.   In addition, the $n$th excited state has $n$
nodes below the real axis (the first excited state has a node at $z=
-0.661296226442715413308i$). The $n$ nodes
 arch above and between the classical turning points at
$E_n^{1/3}e^{-i\pi/6}$ and $E_n^{1/3}e^{-5\pi i/6}$ \cite{Bender:1999ug}.
An interesting question considered in \cite{Bender:2000wj} is what is the precise form of the arch for large $n$?
Unlike for the $N=2$ harmonic oscillator the nodes do not lie on the classical trajectory joining 
two turning points. This trajectory is exactly circular with its centre at the turning point on the imaginary axis \cite{Jennings}.

Our approximations are also useful in computing expectation values. If $\psi(z)$ is an energy
eigenstate 
then the expectation value of $z^m$ can be written as a ratio of contour integrals
\footnote{If $\psi$
is not an energy eigenstate this formula is not
valid; expectation values can be computed via a modified 
inner product including a `charge conjugation' operator \cite{Bender:2004zz}.
The inner product can be related to the standard Dirac inner product via a non-unitary similarity transformation \cite{Mostafazadeh:2002pd}.
}
\begin{equation}
\langle z^m\rangle=
{{\int_C \psi(z)z^m \psi(z)~dz\over{\int_C \psi(z)\psi(z)dz}}},
\end{equation}
 where $C$ is any curve that splits the complex plane into two and starts in one wedge and ends in
 the ${\cal PT}$-symmetric wedge. For $N=3$ one can simply choose $C$ to be the real line:
 \begin{equation}
 \langle z^m\rangle_n={\int^\infty_{-\infty}
 \psi^{n}(x)x^m\psi^{n}(x)~dx
 \over{\int^\infty_{-\infty}\psi^n(x)  \psi^n(x)~dx}},
 \end{equation}
 where $\psi^n$  is the $n$th energy eigenstate ($n=0,1,2,3,...$)
 As the wave functions decay exponentially
these integrals over the real line are well approximated by 
integrals over a finite range $[-\lambda,\lambda]$ for sufficiently large $\lambda$.
We have computed expectation values for the first few energy eigenstates in the $N=3$
model. Here we cut off the integrals at $\lambda=5$ and approximated the $\psi_1$ and $\psi_2$
with the truncation ($p+q\leq 100$) described above.
Plots of the wave functions indicate that  the cut off $\lambda=5$ is good approximation for
the first few eigenstates;  a plot of $\psi^3(x)$ is given in Figure 3.

\begin{figure}	
	\centering
	\begin{subfigure}[t]{2.5in}
		\centering
		\includegraphics[width=2.5in]{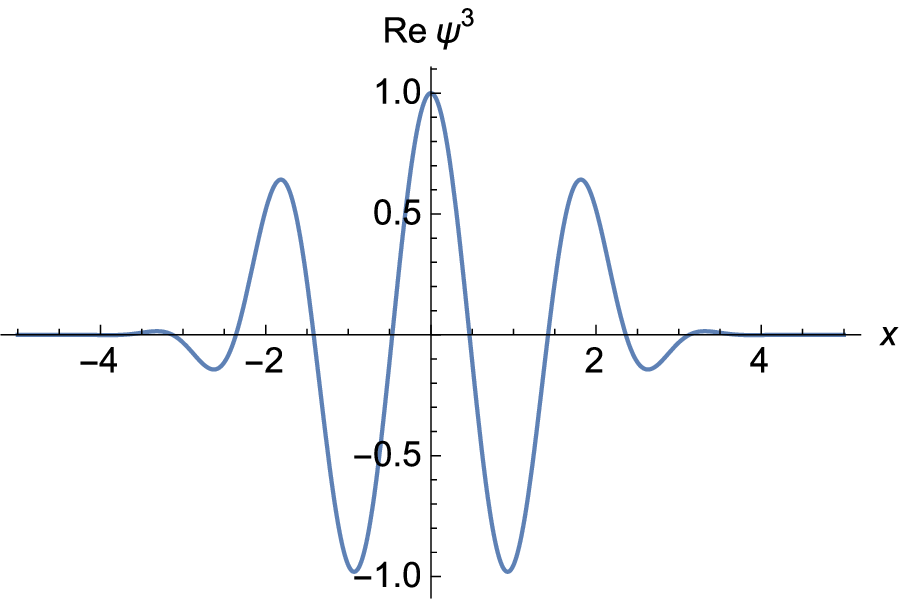}
				
	\end{subfigure}
	\quad
	\begin{subfigure}[t]{2.5in}
		\centering
		\includegraphics[width=2.5in]{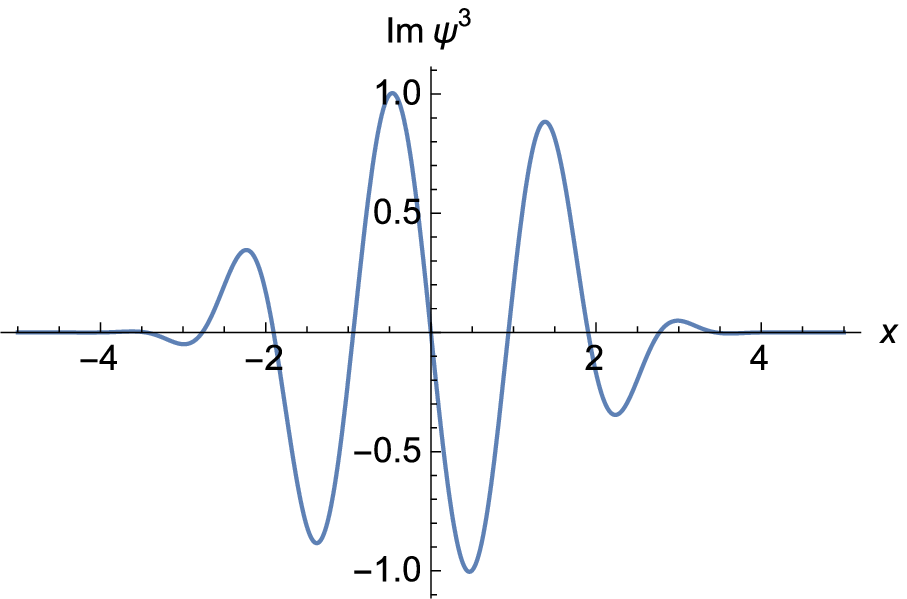}
	
	\end{subfigure}
	\caption{The wave function along the real line for the third excited state.}
	 
	\label{fig:1}
\end{figure}

   \begin{table}[ht]
\caption{Expectation values $\langle z^m \rangle _n$ in the $N=3$ Theory}
\centering
\begin{tabular}{c c c c c  c}
\hline\hline
$~~~~~~n$ & $0$ & $1$ &      $2$& $3$&       \\ % inserts table %heading
\hline
$m$&     & & & &\\
1&-0.5900725330i
&-0.9820718380i 
&-1.2054807539i
&-1.3796870779i\\&     & && \\
3 &   -0.4625068288i 
 &-1.6436915011i 
  & -3.0249095421i    
    &-4.5257687286i\\
4&
-0.3898751086
&  -2.3060330480
      &-5.2092431933
      &-8.9202066199      \\
\hline
\end{tabular}
\label{N=3values}
\end{table}

 Within our approximation $\langle z^2\rangle_m$ is very small (eg. $\langle z^2\rangle_0\approx 10^{-11}$)
 which suggests that $\langle z^2\rangle_n$ is exactly zero. For general $N$
 similar calculations indicate that $\langle z^{N-1}\rangle_n$ is zero
 for all ${\cal PT}$-symmetric pairs of Stokes wedges.
 This is a ${\cal PT}$-symmetric example of Ehrenfest's theorem.
 One can also see a ${\cal PT}$-symmetric virial theorem in the $N=3$ expectation values;
 $\langle z^3\rangle_n$ appears to be exactly $-{2\over5}iE_n$. This can be written as
 $\langle V\rangle_n={2\over5}E_n$. 
 
 \bigskip
 \noindent{\bf Acknowledgements}
 
 \smallskip\noindent
 We are grateful to Carl Bender for useful discussions and correspondence.

\end{document}